\documentclass[aps,twocolumn,showpacs,superscriptaddress,amsmath,amssymb,pre,nofootinbib]{revtex4-1}
\usepackage{epsf,amsmath,amssymb,amsfonts,verbatim,color,multirow,pifont}
\usepackage{graphicx}
\usepackage{dcolumn}% Align table columns on decimal point
\usepackage{bm}% textbf math
\usepackage{txfonts}
\usepackage{hyperref}
\usepackage{soul}

\begin{document}
\title{Exactly solvable two-terminal heat engine with asymmetric Onsager coefficients: Origin of the power-efficiency bound}
\author{Jae Sung Lee}
\author{Jong-Min Park}
\affiliation{School of Physics and Quantum Universe Center, Korea Institute for Advanced Study, Seoul 02455, Korea}
\author{Hyun-Myung Chun}
\affiliation{{Department of Biophysics, University of Michigan, Ann Arbor, Michigan, 48109, USA}}
\author{Jaegon Um}
\affiliation{Department of Physics, Pohang University of Science and Technology, Pohang 37673, Korea}
\author{Hyunggyu~Park}
\email{hgpark@kias.re.kr}
\affiliation{School of Physics and Quantum Universe Center, Korea Institute for Advanced Study, Seoul 02455, Korea}

\newcommand{\revise}[1]{{\color{red}#1}}

\date{\today}

\begin{abstract}
An engine producing a finite power at the ideal (Carnot) efficiency is a dream engine, which is not prohibited by the thermodynamic second law. Some years ago, a two-terminal heat engine with {\em asymmetric} Onsager coefficients in the linear response regime was suggested by Benenti, Saito, and Casati [Phys.~Rev.~Lett.~{\bf 106}, 230602 (2011)], as a prototypical system to make such a dream come true with non-divergent system parameter values. However, such a system has never been realized  in spite of many trials. Here, we introduce an exactly solvable two-terminal Brownian heat engine with the asymmetric Onsager coefficients in the presence of a Lorenz (magnetic) force. Nevertheless, we show that the dream engine regime cannot be accessible even with the asymmetric Onsager coefficients, due to an instability keeping the engine from reaching its steady state. This is consistent with recent trade-off relations between the engine power and efficiency, where the (cyclic) steady-state condition is implicitly presumed. We conclude that the inaccessibility to the dream engine originates from the steady-state constraint on the engine.

\end{abstract}

\pacs{05.70.-a, 05.40.-a, 05.70.Ln, 02.50.-r}

\maketitle

\section{Introduction}
 Is it possible to attain the theoretically maximum efficiency, i.e. the Carnot efficiency $\eta_\textrm{C}$, at a finite power?  As well known from the textbook~\cite{Kittel}, $\eta_\textrm{C}$ is attainable in a reversible or quasi-static process. However, the power of such a reversible engine vanishes as it takes an infinite time to complete one engine cycle. If we operate the engine in a finite-time cycle, we can have a finite power, but usually with irreversible heat dissipation, thus the efficiency should be lower than  $\eta_\textrm{C}$.
 This is why there has been a widespread belief that the {\em dream} engine is impossible, i.e.~it is impossible to achieve $\eta_\textrm{C}$ and a finite power simultaneously, even though there has been no rigorous proof for a long time.

In this context, the recent claim by Benenti, Saito, and Casati (BSC)~\cite{Benenti} was surprising. They showed in the framework of the linear irreversible thermodynamics that the dream engine is possible in a two-terminal thermoelectric device in the presence of a magnetic field breaking the microscopic irreversibility. They considered a thermodynamic system where two currents $J_1$ and $J_2$ are generated by two thermodynamic forces $X_1$ and $X_2$ in the linear response regime as follows:
\begin{align}
J_1 (\textbf{B}) = L_{11} (\textbf{B}) X_1 + L_{12} (\textbf{B}) X_2, \nonumber \\
J_2 (\textbf{B}) = L_{21} (\textbf{B}) X_1 + L_{22} (\textbf{B}) X_2, \label{eq:Onsager}
\end{align}
where $L_{ij}$ is an element of the Onsager matrix ${\textsf L}$ and a function of the magnetic field $\textbf{B}$.
In the case of ${\mathbf B}=0$, the Onsager matrix is proven to be symmetric due to the microreversibility or the detailed balance~\cite{Onsager}. However,
it can be {\em asymmetric} with nonzero ${\mathbf B}$, only satisfying the Onsager-Casimir relation~\cite{Casimir}
as ${\textsf L}({\mathbf B})={\textsf L}^\mathsf{T}(-{\mathbf B})$ with `$\mathsf{T}$' denoting the transpose.
We note that the fluctuation-dissipation relation is still satisfied with non-zero ${\mathbf B}$, while the Onsager
symmetry is broken~\cite{HKLee}.
\begin{figure}
\centering
\includegraphics[width=0.90\linewidth]{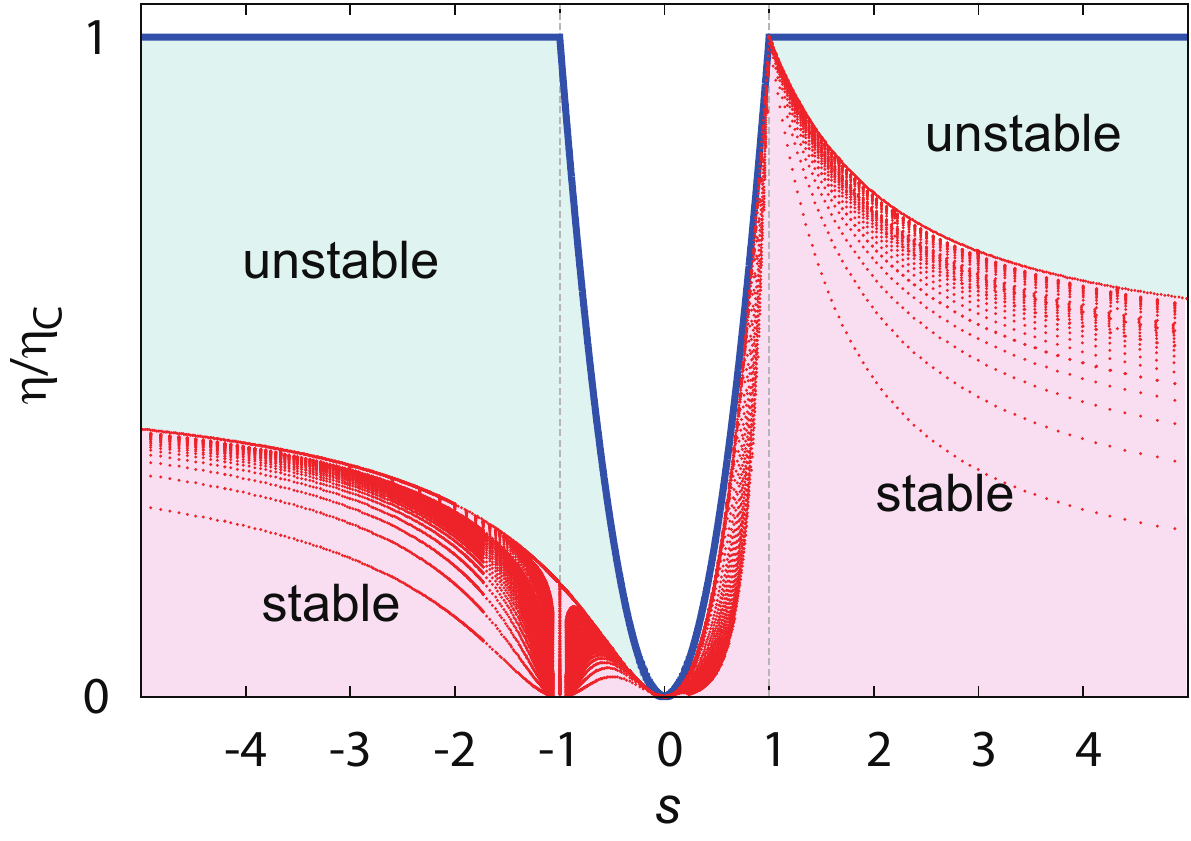}
\caption{Efficiency as a function of $s$. The (blue) solid curve is the maximum efficiency obtained by Benenti \emph{et al}.~\cite{Benenti}, the region below which is allowed by the thermodynamic second law. Scattered (red) points denote the calculated maximum efficiencies of our model at various parameter values subject to the stable steady-state condition. The blueish region above the scattered points is unstable in our model. } \label{fig:stable_region}
\end{figure}
% % % % % % % % % % % % % % % % % % % % %

BSC~\cite{Benenti} showed that Carnot efficiency at a finite power is attainable when the following conditions are satisfied:
\begin{align}
\mathcal{L} \equiv 4\det{\textsf{L}}  - (L_{12}-L_{21})^2 = 0~~~\textrm{and}~~\left|s\equiv \frac{L_{12}}{L_{21}}\right|>1, \label{eq:L}
\end{align}
where `$\det$' denotes the determinant  and $s$ is called the symmetry factor. The first equation represents the maximum efficiency condition for given $s$.
This result is presented in Fig.~\ref{fig:stable_region} as the solid curve, which is the curve of the maximum efficiency as a function of $s$ constrained by the thermodynamic second law. One can see that $\eta_\textrm{C}$ is accessible for $|s|\ge 1$, where the power (proportional to $s^2-1$), is finite except for the symmetric case ($s=1$). This suggests that the dream engine could be possible with
a symmetry breaking induced by the magnetic field.

This study triggered a flurry of subsequent discussions on developing engine mechanisms achieving the Carnot efficiency at a finite power or in an irreversible process~\cite{Brandner, Allahverdyan1, Karel, Campisi, Shiraish, Polettini, Holubec1, Andreas, Pietzonka, JSLee1, JSLee2}. From these studies, several mechanisms have been suggested to realize the dream engine, for example, by approaching the criticality of the engine system~\cite{Campisi}, infinitely fast process~\cite{Polettini}, and cycling in the diverging damping coefficient (or vanishing-relaxation-time) limit~\cite{Holubec1}.
More importantly, several trade-off relations between the power and the efficiency have been found
for various situations~\cite{Shiraish, Andreas, Pietzonka}  such as
\begin{align}
\mathcal{P} \leq \Theta (\eta_\textrm{C} - \eta), \label{eq:tradeoff}
\end{align}
where $\mathcal{P}$ is the power, $\eta$ is the efficiency, and $\Theta$ is a system-dependent positive constant. This relation sets a constraint that the power should vanish to attain $\eta_\textrm{C}$ unless $\Theta$ diverges. All these findings strongly assert that some diverging limits are necessary to attain the dream engine.

On the other hand, the BSC formulation~\cite{Benenti} does not require any divergence of parameters for achieving $\eta_\textrm{C}$ at a finite power. In other words, if we have the model described by Eq.~\eqref{eq:Onsager} with $s\neq 1$ and find a set of parameters with moderate values satisfying Eq.~\eqref{eq:L}, the dream engine should be realized. In this sense, the BSC theory~\cite{Benenti} and all the subsequent studies look contradictory. Therefore, it is important to study a concrete two-terminal model with asymmetric Onsager coefficients for investigating the possibility attaining the Carnot efficiency at a finite power in a realistic situation with moderate parameters.

However, nobody has succeeded in finding such a two-terminal engine with $s\neq 1$. In a purely coherent two-terminal system, for example, the off-diagonal elements of the Onsager matrix turn out to be even functions of the magnetic field, thus, they are always symmetric and no {\em reversible} currents responsible for the dream engine are possible~\cite{Buttiker, Brandner}. Inelastic scatterings and interactions are suggested to break the symmetry, but no explicit cases are reported.
To detour this problem, some studies introduced a third terminal (or more terminals) with a specific condition for mimicking a two-terminal engine~\cite{Brandner,Brandner2}, a time-averaged Onsager matrix for a periodically driven system~\cite{Karel}, and the Nernst effect~\cite{Stark}. However, they are not exactly matched to the two-terminal system described by Eq.~\eqref{eq:Onsager} and no dream engine was realized in the steady state.
%\textcolor{red}{Due to absence of the two-terminal system with the broken symmetry, it has been regarded that such a system is not allowed in nature.}

In this study, we introduce an exactly solvable stochastic model which manifests the symmetry breaking of the Onsager matrix in the presence of a magnetic field. We find that many sets of parameters with moderate values satisfy the dream engine condition in Eq.~\eqref{eq:L}. Nevertheless, this does not guarantee the existence of the dream engine alone, because one should check the stability of the steady
state for such a set of parameters. It turns out that there is no stable steady state in all those sets of parameters satisfying the dream engine condition. Our finding stresses the importance of the \emph{boundary condition} or \emph{intrinsic constraint} imposed for an engine problem, which is the steady-state or periodic-cycle condition, inevitably required for steady production of work from an engine.
We conclude that this constraint plays the most crucial role in forbidding the dream engine realized, rather than the symmetry breaking of the Onsager matrix which is a necessary condition.

\section{Model}

\begin{figure}
\centering
\includegraphics[width=0.85\linewidth]{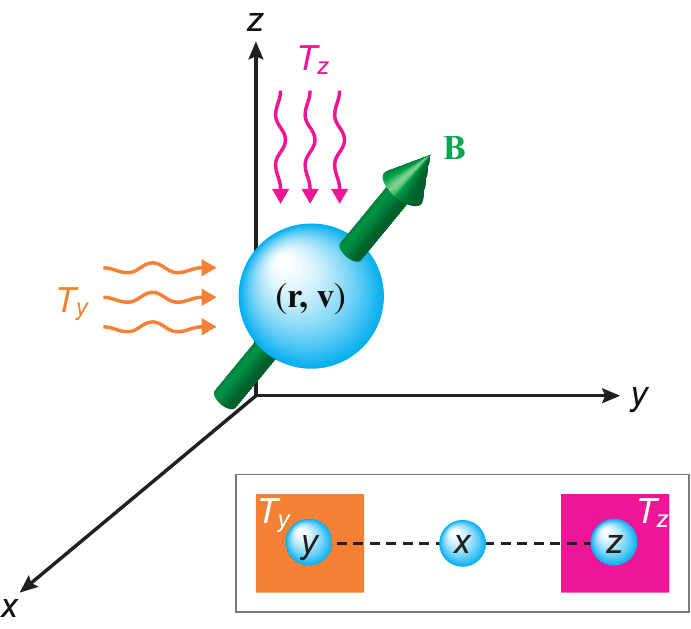}
\caption{The two-terminal Brownian engine in the three-dimensional space. (Inset) This model can be interpreted as a three-particle
system in the one-dimensional space with one particle outside of the heat reservoirs. } \label{fig:model}
\end{figure}

 We consider an underdamped Brownian dynamics of a charged particle with mass $m$ in the three-dimensional space as illustrated in Fig.~\ref{fig:model}. Its position and velocity are denoted by $\textbf{r}=(x,y,z)^{\textsf{T}}$ and $\textbf{v}=(v_x,v_y,v_z)^{\textsf{T}}$, respectively. The particle moves in a magnetic field $\textbf{B}=(B_x, B_y, B_z)^{\textsf{T}}$ and is confined in a harmonic potential with stiffness $k(>0)$. Its dynamics along the $y$- and $z$-axis are affected by heat reservoirs with different temperatures $T_y$ and $T_z$, respectively, while the dynamics along the $x$-axis is not affected by any heat reservoir, thus, deterministic~\cite{exp1}.
 A linear external nonconservative force (torque), $\textbf{f}_\textrm{nc}=\epsilon y \hat{x} +\delta x \hat{y}$, is applied to extract work out of the engine.

% % % % % % % % % % % % % % % % % %
% % % % % % % % % % % % % % % % % % % % %

 The Langevin equation for this particle can be written as
\begin{align}
&\textbf{v} = \dot{\textbf{r}},~ m \dot{\textbf{v}}= - k \textbf{r} +\textsf{F}_\textrm{nc} \textbf{r} + \textbf{v} \times \textbf{B} - \Gamma \textbf{v} +\boldsymbol{\xi}  ~~~\textrm{with} \label{eq:main_eq} \\
&\textsf{F}_\textrm{nc}= \left( \begin{array}{ccc}
0 & \epsilon & 0 \\
\delta & 0 & 0 \\
0 & 0 & 0 \\
\end{array} \right),
\Gamma = \left( \begin{array}{ccc}
0 & 0 & 0 \\
0 & \gamma & 0 \\
0 & 0 & \gamma \\
\end{array} \right), \textrm{ and }~
\boldsymbol{\xi} =  \left( \begin{array}{c}
0  \\
\xi_y  \\
\xi_z \\
\end{array} \right), \nonumber
\end{align}
where $\xi_i$ $(i=y,z)$ is a white Gaussian noise satisfying $\langle \xi_i (t) \xi_j  (t^\prime) \rangle =2\gamma T_i \delta_{ij} \delta(t-t^\prime)$ in the Boltzmann unit $(k_\textrm{B} =1)$ and
$\textbf{v} \times \textbf{B}$ is the Lorentz force. Note that $\textbf{f}_\textrm{nc}(=\textsf{F}_\textrm{nc} \textbf{r})$  becomes conservative  when $\epsilon = \delta$, otherwise nonconservative, then drives the system out of equilibrium.  In addition, the temperature difference between $T_y$ and $T_z$  is another driving force. Thus, there are two thermodynamic forces driving the system into a nonequilibrium state such as
\begin{align}
X_1 \equiv\delta-\epsilon~~~\textrm{and}~~~X_2\equiv 1/T_y - 1/T_z~~(T_y<T_z)~. \label{eq:force}
\end{align}
Note that the Carnot efficiency is given as $\eta_\textrm{C}=T_y X_2$.

The two-dimensional version has been studied in various contexts with and without a magnetic field~\cite{Crisanti,ParkJM,Filliger,Chiang,LeeKwon,Chun1}
and the Onsager coefficients turn out to be symmetric even in the presence of a magnetic field (see Appendix A). This is why we resort to a more complicated three-dimensional version, still keeping only two terminals. Equation~\eqref{eq:main_eq} can be also interpreted as a three-particle system in the one-dimensional space, each of which is confined in a harmonic potential and interacts to each other through $\textbf{f}_\textrm{nc}$
and the Lorenz force as illustrated in Inset of Fig.~\ref{fig:model}. Two particles are in contact with two different heat reservoirs, respectively and the remaining one particle is outside of the reservoirs. The two-dimensional version does not carry this extra particle
with the $y$-$z$ exchange (left-right) symmetry.

In our model, we calculate the heat transferred from the $i$-axis reservoir into the particle $Q_i (t)$ and the work extraction due to the nonconservative force $W(t)$ by the standard stochastic energetics~\cite{ParkJM,Sekimoto}.  During an infinitesimal time interval $[t,t+dt]$, their incrementals can be written as \begin{align}
&dQ_i (t)= v_i (t) \circ [-\gamma v_i (t) dt + d\Xi_i (t)]  , \\
&dW(t) = -\textbf{f}_\textrm{nc} \cdot d\textbf{r} = -[\epsilon v_x (t) y(t)+\delta x(t) v_y(t)]dt,
\end{align}
where $\circ$ denotes the Stratonovich multiplication~\cite{Sekimoto} and  $d\Xi_i (t) \equiv \int_t^{t+dt} dt^\prime \xi_i (t^\prime)$ satisfying $\langle  d\Xi_i (t)\rangle =0$ and $\langle  d\Xi_i (t) d\Xi_j (t)\rangle =2\gamma T_i \delta_{ij} dt$. From the thermodynamic first law, $dE(t) = dQ_y (t) +dQ_z (t) - dW(t) $, where $dE(t)$ is the internal energy change during $[t,t+dt]$.
We consider the steady state average only, denoted by $\langle \cdots \rangle_\textrm{s}$. As  $\langle dE \rangle_\textrm{s}=0$, we have two independent energy currents. From the Stratonovich algebra, $\langle v_i \circ d\Xi_i (t) \rangle_\textrm{s} = \gamma T_i dt/m$, the rates of the heat and work are given by
\begin{align}
q_i &\equiv  \langle \dot{Q}_i \rangle_\textrm{s}= \frac{\gamma}{m} (T_i - m\langle v_i^2  \rangle_\textrm{s}), \label{eq:heat_rate}\\
\mathcal{P} &\equiv  \langle P \rangle_\textrm{s} = (\epsilon-\delta)  \langle x v_y \rangle_\textrm{s}. \label{eq:work_rate}
\end{align}
where $\dot{Q_i} = dQ_i/dt$, $P = dW/dt$, and the second equation is obtained by using the steady-state property as $\frac{d}{dt} \langle x(t) y(t) \rangle_\textrm{s} = \langle v_x (t) y(t) \rangle_\textrm{s} + \langle x(t) v_y(t) \rangle_\textrm{s} =0$.

\section{Onsager coefficients}

We define two currents $J_1$ and $J_2$ as follows:
\begin{align}
J_1 \equiv \frac{\langle x v_y \rangle_\textrm{s}}{T_y},~J_2  \equiv q_z~, \label{eq:J1J2}
\end{align}
where $q_z$ is the heat current out of the high-temperature reservoir  and
the work current (power) is given by $\mathcal{P}=-J_1X_1T_y$, as in the standard linear irreversible thermodynamics~\cite{Onsager}.
Then, the total entropy production (EP) rate $\langle\dot{S}_\textrm{tot}\rangle_\textrm{s}$ can be written as
\begin{align}
\langle \dot{S}_\textrm{tot} \rangle_\textrm{s} = -\frac{q_y}{T_y} - \frac{q_z}{T_z} = J_1 X_1 + J_2 X_2~, \label{eq:entropy_rate}
\end{align}
and the thermodynamic second law puts a constraint on the Onsager matrix as
\begin{align}
\mathcal{L} =4\det{\textsf{L}}  - (L_{12}-L_{21})^2 \ge 0 ~~\textrm{for}~~ L_{11}, L_{22} >0~.\label{eq:2ndlaw}
\end{align}
Note that, in the so-called {\em tight-coupling} case with $\det {\textsf{L}}=0$~\cite{vdBroeck},
the Onsager symmetry $(s=1)$ is required by the above constraint.

We now calculate $J_1$ and $J_2$ explicitly, i.e.~$\langle x v_y \rangle_\textrm{s}$ and $\langle v_z^2 \rangle_\textrm{s}$ by following the standard procedure for solving a multivariate Ornstein-Uhlenbeck process~\cite{Gardiner,LeePark2}.
Introduce a state vector $\textbf{z} \equiv (x,y,z,v_x,v_y,v_z)^{\textsf{T}}$ and a noise vector $d \Xi(t) \equiv (d\Xi_1(t),  d\Xi_2 (t), \cdots, d\Xi_6 (t))^{\textsf{T}}$ with $\langle d\Xi (t)\rangle =0$ and $\langle d\Xi(t) d\Xi^{\textsf{T}} (t)   \rangle = 2\textsf{D} dt$, where $\textsf{D}$ is a $6 \times 6$ symmetric diffusion matrix.
Then, the equation of motion, Eq.~\eqref{eq:main_eq}, can be written in the form of the Ornstein-Uhlenbeck process as
\begin{align}
d\textbf{z} = - \textsf{A} \textbf{z} dt +  d\Xi \label{eqA:OU},
\end{align}
where
\begin{align}
\textsf{A}= \frac{1}{m} \left( \begin{array}{cccccc}
0 & 0 & 0 & -m & 0 & 0 \\
0 & 0 & 0 & 0 & -m & 0 \\
0 & 0 & 0 & 0 & 0 & -m \\
k & -\epsilon & 0 & 0 & -B_z & B_y \\
-\delta & k & 0 & B_z & \gamma & -B_x \\
0 & 0 & k & -B_y & B_x & \gamma \\
\end{array} \right).
\end{align}
and $\textsf{D}_{ij}=0$ for all elements except $\textsf{D}_{55} = {\gamma T_y}/m^2$ and $\textsf{D}_{66} = {\gamma T_z}/m^2$.

The covariant matrix $\Sigma$ is defined as $\Sigma \equiv \langle \textbf{z} \textbf{z}^{\textsf{T}} \rangle_\textrm{s}=\Sigma^{\textsf{T}}$,
which satisfies
\begin{align}
\textsf{A} \Sigma + \Sigma \textsf{A}^{\textsf{T}} = 2 \textsf{D}~ \label{eqA:OUsln}
\end{align}
from the steady-state condition $d\Sigma=0$~\cite{Gardiner,LeePark2}.
It is straightforward to solve Eq.~\eqref{eqA:OUsln} in general, but its solution for $\Sigma$ is quite complicated.
In order to calculate the Onsager coefficients in Eq.~\eqref{eq:Onsager}, it is convenient to employ a perturbation expansion
near the steady state (equilibrium) when $\delta=\epsilon$  and $T_z=T_y$, instead.
Up to the lowest order in the thermodynamic forces $X_1$ and $X_2$ in Eq.~\eqref{eq:force}, we expand the matrices as
\begin{align}
&\textsf{A}=\textsf{A}_0+\textsf{A}_1 X_1~,~~\textsf{D}=\textsf{D}_0 + \textsf{D}_2 X_2~, \nonumber\\
&{\Sigma}=\Sigma_0+\Sigma_1 X_1 + \Sigma_2 X_2 , \label{eq:OUper1}
\end{align}
where the unperturbed ones
$\textsf{A}_0= \textsf{A}|_{\delta=\epsilon}$ and  $\textsf{D}_0=\textsf{D}|_{T_z = T_y}$, and
the fist-order corrections
$[\textsf{A}_1]_{ij}=0$ except $[\textsf{A}_1]_{51}=-1/m$ and
$[\textsf{D}_2]_{ij}=0$ except $[\textsf{D}_2]_{66}=\gamma T_y^2/m^2$ for all $i$ and $j$.

The covariant matrix expansion with $\Sigma_0$, $\Sigma_1$, and $\Sigma_2$, can be obtained by a series of equations
derived from Eq.~\eqref{eqA:OUsln} as
\begin{align}
&\textsf{A}_0 \Sigma_0 + \Sigma_0 \textsf{A}_0^{\textsf{T}} = 2 \textsf{D}_0~, \nonumber \\
&\textsf{A}_0 \Sigma_1 + \Sigma_1 \textsf{A}_0^{\textsf{T}} = -\textsf{A}_1 \Sigma_0 - \Sigma_0 \textsf{A}_1^{\textsf{T}}~,\label{eq:OUper2}\\
&\textsf{A}_0 \Sigma_2 + \Sigma_2 \textsf{A}_0^{\textsf{T}} = 2 \textsf{D}_2~.\nonumber
\end{align}
First, we find
\begin{align}
\Sigma_0=  T_y \left( \begin{array}{cccccc}
k/K & \epsilon/K & 0 & 0 & 0 & 0 \\
\epsilon/K & k/K & 0 & 0 & 0 & 0 \\
0 & 0 & 1/k & 0 & 0 & 0 \\
0 & 0 & 0 & 1/m & 0 & 0 \\
0 & 0 & 0 & 0 & 1/m& 0 \\
0 & 0 & 0 & 0 & 0 & 1/m\\
\end{array} \right)~,\label{eq:Sigma0}
\end{align}
with $K=k^2-\epsilon^2$. The stability of the unperturbed steady state is guaranteed by the positivity of all eigenvalues~\cite{Gardiner,LeePark2}, which gives
\begin{align}
K=k^2-\epsilon^2 >0~~\textrm{(stability condition)}~.\label{eq:stability}
\end{align}
We can also find $\Sigma_1$ and $\Sigma_2$ from Eqs.~\eqref{eq:OUper2} and \eqref{eq:Sigma0}.

From Eqs.~\eqref{eq:Onsager} and \eqref{eq:J1J2}, we express the Onsager matrix $\textsf{L}$ by
the elements of the covariant matrix $\Sigma$ as
\begin{align}
\textsf{L}=  \left( \begin{array}{cc}
[\Sigma_1]_{15}/T_y & [\Sigma_2]_{15}/T_y  \\
-\gamma [\Sigma_1]_{66} & \gamma \left(T_y^2/m-[\Sigma_2]_{66}\right) \\
\end{array} \right)~.\label{eq:OnsagerL}
\end{align}
For simplicity, we set $B_x=0$ as an example. Then we get
%\begin{widetext}
\begin{align}
L_{11} &= \frac{1}{\gamma\mathcal{G}}
\left[ (2k^2-\epsilon^2)\gamma^2C_0  C_2  +k B_z^2 (C_1+m\epsilon^2)C_2 \right. \nonumber\\
&\qquad\qquad\qquad\qquad \left. +   m\epsilon^2( m \epsilon^2 B_z^2+2 k\gamma^2  B_y^2) \right]~, \nonumber\\%\label{eq:L11} \\
L_{22} &= \frac{\gamma T_y^2 B_y^2 }{m\mathcal{G}}
\left[(2k^2-\epsilon^2)(2\gamma^2 C_3+m\epsilon^2B_y^2)+2k(C_3^2+k^2B_y^2B_z^2) \right]~, \nonumber\\%\label{eq:L22} \\
L_{12} &= \frac{\epsilon T_y B_y^2 }{\mathcal{G}} \left[ (2k^2-\epsilon^2)\gamma C_2+2k\gamma m\epsilon^2
 -\epsilon  B_z C_1 \right]~, \nonumber \\
L_{21} &= \frac{\epsilon T_y B_y^2 }{\mathcal{G}} \left[ (2k^2-\epsilon^2)\gamma C_2+2k\gamma m\epsilon^2 +\epsilon  B_z C_1\right]~,  \label{eq:L12}
\end{align}
where $C_0$, $C_1$, $C_2$, $C_3$, and $\mathcal{G}$ are given as
\begin{align}
&C_0 = B_y^2 +B_z^2, C_1 = k C_0 + m \epsilon^2, C_2= C_0 + 2\gamma^2, C_3 = k B_z^2 + m \epsilon^2,\nonumber \\
%C_3 &= k B_z^2 + m \epsilon^2, \textrm{~and~}
&\mathcal{G} = \left[\{(2k^2-\epsilon^2)\gamma^2+k(C_1+m\epsilon^2)\} C_2 +(m\epsilon^2)^2\right] C_1 . \label{eq:constants}
\end{align}
Note that all $C_i$'s ($i=0,1,2,3$) are positive and the even functions of $B_y$ and $B_z$. The odd function in terms of the magnetic
field appears only in the last term of the off-diagonal elements, $L_{12}$ and $L_{21}$.

As expected, the Onsager-Casimir relation~\cite{Onsager,Casimir} is satisfied as ${\mathbf L}({\mathbf B})={\mathbf L}^\mathsf{T}(-{\mathbf B})$, but the Onsager symmetry
is broken; ${\mathbf L}({\mathbf B})\neq {\mathbf L}^\mathsf{T}({\mathbf B})$, seen in Eq.~\eqref{eq:L12}.
In contrast to the two-dimensional case, we find indeed a two-terminal model with the asymmetric Onsager matrix, i.e. $s\neq 1$ for
the three-dimensional version.

It is interesting to note that the Onsager matrix becomes symmetric ($s= 1$) when $B_z=0$ with $B_y\neq 0$ in Eq.~\eqref{eq:L12}. Moreover,
$\mathcal{L}=4 \textrm{det} (\textsf{L})=0$ (tight-coupling), implying that the reversible process is possible with
$\langle \dot{S}_\textrm{tot}\rangle_\textrm{s}=0$ in Eq.~\eqref{eq:entropy_rate} at $X_1=-\epsilon T_y X_2$ and thus the efficiency $\eta$
can reach the Carnot efficiency $\eta_\textrm{C}$.

\section{Efficiency, Power, and EP rate}

The engine efficiency $\eta$ in converting the heat flowing from the high temperature reservoir into the power is defined as
\begin{align}
\eta=\frac{\mathcal{P}}{q_z}=\frac{-J_1X_1T_y}{J_2}=\frac{-T_y X_1 (L_{11}X_1+L_{12}X_2)}{L_{21}X_1+L_{22}X_2}~,\label{eq:eta}
\end{align}
which is maximized for a given temperature gradient $X_2$ at
\begin{align}
X_1=X_1^*=-\frac{L_{22}}{L_{21}}  \left(1 -\sqrt{ \frac{\det {\textsf{L}} }{L_{11}L_{22} }} \right) X_2 \label{eq:X1X2}
\end{align}
with the maximum efficiency for given $\textsf{L}$
\begin{align}
\eta^*=\eta(X_1^*)=\eta_\textrm{C}\frac{L_{11}L_{22}}{L_{21}^2} \left(1 -\sqrt{ \frac{\det {\textsf{L}} }{L_{11}L_{22} }} \right)^2~. \label{eq:eta^*}
\end{align}
where $X_2$ is replaced by $\eta_\textrm{C}=T_yX_2$.

It is rather convenient to rewrite $\eta^*$ in terms of $\mathcal{L}$ in Eq.~\eqref{eq:2ndlaw} as
\begin{align}
\eta^*=\frac{\eta_\textrm{C}}{4} \left[\sqrt{\mathcal{Y}+(s+1)^2}-\sqrt{\mathcal{Y}+(s-1)^2}\right]^2~~(\mathcal{Y}=\mathcal{L}/L_{21}^2), \label{eq:eta^*Y}
\end{align}
with $\mathcal{Y}\ge 0$ by the thermodynamic constraint in Eq.~\eqref{eq:2ndlaw}.
One can easily find that $\eta^*$ is a monotonically decreasing function of $\mathcal{Y}$ for fixed $s$, so $\eta^*$ can
reach its highest value $\eta^\textrm{max}$ at $\mathcal{Y}=0$ as
\begin{align}
\eta^\textrm{max}= \left\{\begin{array}{ll}
\eta_\textrm{C}  & \textrm{for}~ |s|\ge 1 \\
s^2\eta_\textrm{C}  & \textrm{for}~ |s|<1 \\
\end{array} \right.~, \label{eq:eta^max}
\end{align}
which is shown as the blue solid curve in Fig.~\ref{fig:stable_region}~\cite{Benenti}.
Note that, in the symmetric case ($s=1$),
the Carnot efficiency is achieved  in the tight-coupling limit ($\det{\textsf{L}}= 0$).

The power and the EP rate at the maximum efficiency $\eta^*$ are given as
\begin{align}
\mathcal{P}^*&=\mathcal{P}(X_1^*)=L_{22}\sqrt{ \frac{\mathcal{Y}+(s-1)^2}{\mathcal{Y}+(s+1)^2}}~\eta^*  X_2~, \label{eq:power}\\
\langle \dot{S}_\textrm{tot} \rangle_\textrm{s}^*&=L_{22}\sqrt{\frac{\mathcal{Y}+(s-1)^2}{\mathcal{Y}+(s+1)^2} } \left(1-\frac{\eta^*}{\eta_\textrm{C}}\right) X_2^2~.\label{eq:EP}
\end{align}
Along the highest efficiency curve in Eq.~\eqref{eq:eta^max}, the power $\mathcal{P}^\textrm{m}$ and the EP rate
$\langle \dot{S}_\textrm{tot} \rangle_\textrm{s}^\textrm{m}$ are obtained as
\begin{align}
\mathcal{P}^\textrm{m}&= \frac{L_{22}\eta^2_\textrm{C}}{T_y}
\left\{\begin{array}{ll}
\left|\frac{s-1}{s+1}\right|  & \textrm{for}~ |s|\ge 1 \\
s^2\left(\frac{1-s}{1+s}\right)   & \textrm{for}~ |s|<1 \\
\end{array} \right.~\label{eq:powerm}\\
\langle \dot{S}_\textrm{tot} \rangle_\textrm{s}^\textrm{m}&=
\frac{L_{22}\eta^2_\textrm{C}}{T_y^2}
\left\{\begin{array}{ll}
0  & \textrm{for}~ |s|\ge 1 \\
(1-s)^2   & \textrm{for}~ |s|<1 \\
\end{array} \right.~.\label{eq:EPm}
\end{align}
For $|s|> 1$, we find that the efficiency can reach $\eta_\textrm{C}$ in Eq.~\eqref{eq:eta^max} with nonzero power $\mathcal{P}^\textrm{m}$ in Eq.~\eqref{eq:powerm}  (dream engine) and vanishing EP in Eq.~\eqref{eq:EPm}, which was the main result of BSC~\cite{Benenti}.

\section{stability}
As in Eqs.~\eqref{eq:Sigma0} and \eqref{eq:stability}, the unperturbed steady state (equilibrium) is stable only for $K=k^2-\epsilon^2>0$.
Thus, we should examine the results of the last section within the stability condition. It is easy to see that
$L_{11}, L_{22} >0$ and $\mathcal{G}>0$ for $k^2>\epsilon^2$ in Eqs.~\eqref{eq:L12} and \eqref{eq:constants}.
We need to check whether the dream engine condition, i.e.~$\mathcal{L}=0$ for $|s|>1$ in Eq.~\eqref{eq:L}, is possible for $k^2>\epsilon^2$.

We rewrite $\mathcal{L}$ explicitly, using Eq.~\eqref{eq:L12}, as
\begin{align}
\mathcal{L}=\frac{4T_y^2 B_y^2}{m\mathcal{G}^2}\left\{[l_{11}][l_{22}]-m\epsilon^2 B_y^2 \gamma^2 [(2k^2-\epsilon^2)C_2 +2km\epsilon^2]^2
\right\}~, \label{eq:L1}
\end{align}
where $[l_{11}]$ and $[l_{22}]$ are the expressions inside of the $[\cdot\cdot]$ of $L_{11}$ and $L_{22}$, respectively,  in Eq.~\eqref{eq:L12}. First, $\mathcal{L}=0$ and $s=1$ for $B_z=0$ pointed out in Sec.~III. Second, $\mathcal{L}$ is the even function of $B_z$. Thus, $\mathcal{L}$ can be written in a power series of $B_z^2$ as
\begin{align}
\mathcal{L}=\frac{4T_y^2 B_y^2}{m\mathcal{G}^2}\sum_{n=1}^5 a_{2n} B_z^{2n}~,
\end{align}
where the coefficient $a_{2n}$ is a function of $m$, $k$, $\epsilon^2$, $\gamma^2$, and $B_y^2$.

\begin{figure}
\centering
\includegraphics[width=0.90\linewidth]{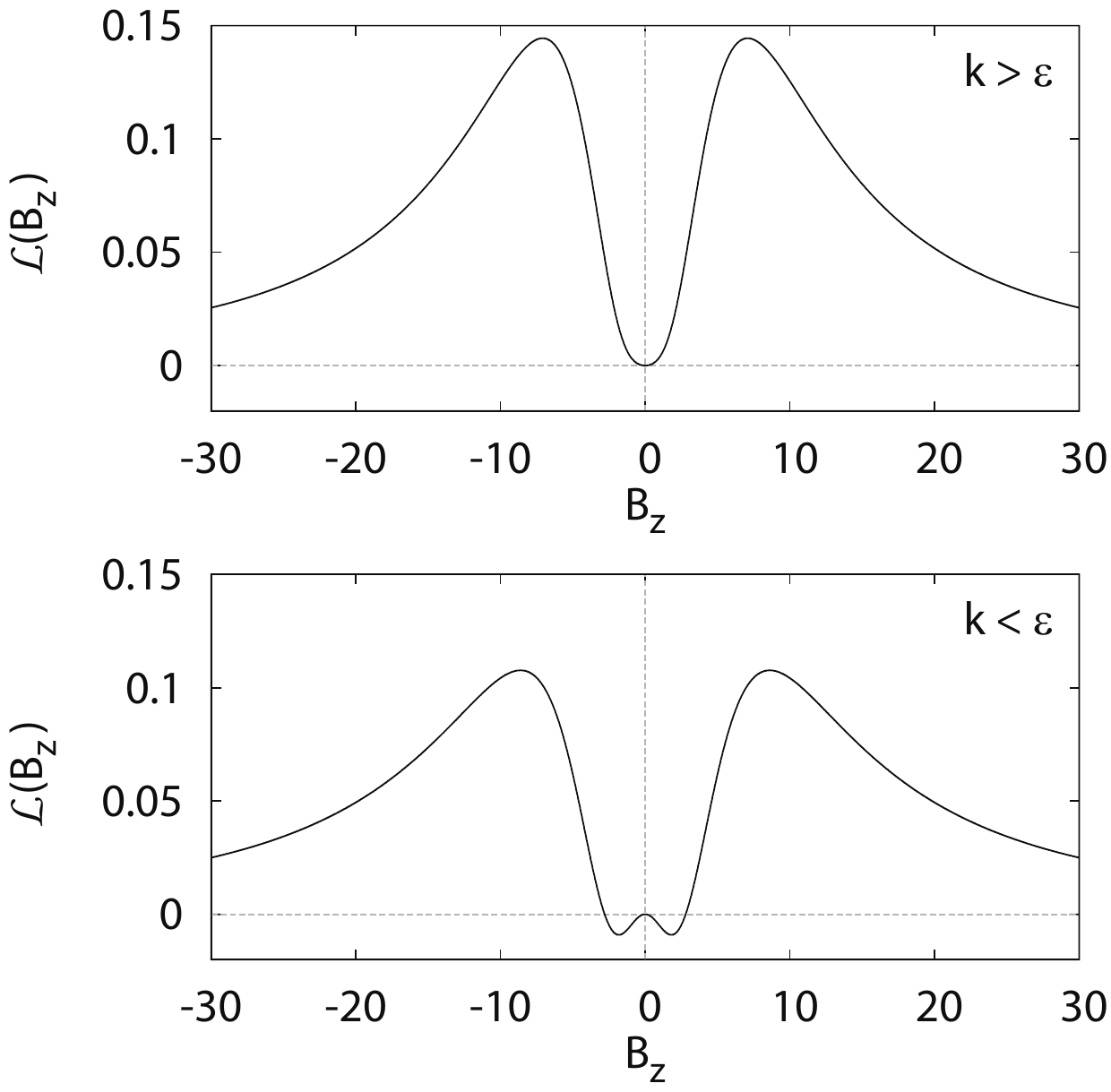}
\caption{  Plots of $\mathcal{L}$ as a function of $B_z$  when (a) $k >|\epsilon|$ and (b) $0<k <|\epsilon|$, respectively. For both cases, $\mathcal{L} = 0$ at $B_z =0$ and $\mathcal{L}\rightarrow 0^+$ in $|B_z| \rightarrow \infty$ limit. } \label{fig:Lfunction}
\end{figure}

It is straightforward to prove that all coefficients $a_{2n}$'s are definitely positive for $k^2>\epsilon^2$ (not shown here), which implies
that the $\mathcal{L}=0$ condition is satisfied only at $B_z=0$, thus $s=1$.  Thus, the dream engine
can not be achieved for any set of parameters compatible with the stability condition. In Fig.~~\ref{fig:Lfunction},
$\mathcal{L}$ versus $B_z$ is plotted for a typical parameter set when (a) $k> |\epsilon|$  and  (b) $0<k<|\epsilon|$.
Note that $\mathcal{L}$ can vanish at a nonzero $B_z$ only in the unstable case (b).
Our result for this exactly solvable model clearly shows the key role of the intrinsically imposed constraint, i.e.~the existence of a stable steady state in an engine problem.

% % % % % % % % % % % % % % % % % %
% % % % % % % % % % % % % % % % % % % % %

We numerically check the maximum efficiency values in the stable region.
As $\eta^*$ is the monotonically decreasing function of $\mathcal{Y}$ for a given $s$ in Eq.~\eqref{eq:eta^*Y}, the highest possible efficiency value
can be obtained at the smallest possible $\mathcal{Y}$, subject to the stability condition ($k>|\epsilon|$).

For this calculation, we vary $k$ ($0\leq k \leq 7$) , $B_z$ ($-2500 \leq B_z \leq 2500$), $m$ ($4 \leq m \leq 10^6$), $\gamma$ ($0.01 \leq \gamma \leq 1$), $100 \leq B_y \leq 10^6$, and $1 \leq T_2 \leq 10^6$  with fixed parameter $B_x=0$. The results are presented in Fig.~\ref{fig:stable_region}, where
the stable region does not reach the Carnot efficiency line except $s=1$. Note that the stable region is much smaller for negative $s$
and in particular does not exist for $s=-1$.  This is special in our model with $B_x=0$, which can be easily noticed in Eq.~\eqref{eq:L12},
i.e., $L_{12}+L_{21} \propto (2k^2-\epsilon^2)C_1 + 2km\epsilon^2$ can never be zero for $k^2>\epsilon^2$.

\section{Summary and Discussion}

In summary, we explicitly showed in an exactly solvable model that the stability constraint for the steady state is crucial in prohibiting
a dream engine. The asymmetry of the Onsager matrix $\textsf{L}$ may arise in a two-terminal engine, but
the reversible limit for a dream engine can not be accessible due to the stability condition of the unperturbed steady state.

The power-efficiency trade-off relation derived by Dechant and Sasa (DS)~\cite{Andreas} should be applied to our model, which includes
a non-conservative force in the framework of an underdamped dynamics. The DS derivation is based on the entropic bound on general irreversible currents, which is written as
\begin{align}
\langle \dot{Q}_i \rangle^2 \leq \zeta_i \langle \dot{S}_\textrm{tot}\rangle~,
\label{eq:dechant-sasa}
\end{align}
where
%$\zeta(t)$ is a time-dependent constant depending on a given system and
$\langle \cdots \rangle$ denotes the ensemble average at an arbitrary time $t$,
a time-dependent coefficient $\zeta_i=\gamma T_i \langle v_i^2\rangle$, and
$\langle \dot{S}_\textrm{tot} \rangle= \langle \dot{S}_\textrm{sys} \rangle - \langle\dot{Q}_y \rangle/T_y- \langle \dot{Q}_z \rangle /T_z$  with the Shannon entropy change rate  $\langle\dot{S}_\textrm{sys}\rangle$. Note that this entropic bound is valid even  with the
Lorentz force. Then, we can show that the instantaneous power
\begin{align}
\langle P \rangle \leq \frac{\zeta_z\eta}{T_y} \left[\eta_\textrm{C} -\eta + \frac{T_y\langle \dot{S}_\textrm{sys} \rangle - \langle \dot{E} \rangle  }{\langle \dot{Q}_z \rangle}  \right]~,
\label{eq:dechant-sasa1}
\end{align}
where $\langle \dot{E} \rangle$ is the system-energy change rate.
In the steady state with $\langle \dot{S}_\textrm{sys} \rangle=\langle \dot{E} \rangle=0$, Eq.~\eqref{eq:dechant-sasa1} returns back to Eq.~\eqref{eq:tradeoff}.
If the system is in a transient state, the power may not vanish at $\eta = \eta_\textrm{C}$ in general.  This clearly shows the importance of the steady-state constraint for the power-efficiency bound.
The above discussion can be extended to a cyclic engine. The similar bound as in Eq.~\eqref{eq:tradeoff} can be derived in a cyclic steady state~\cite{Andreas}, where the Shannon entropy change of the system over one cycle is zero. In the Appendix B, the detailed derivation for the work extraction per cycle is given for a cyclic engine.

In conclusion, we show that the steady-state constraint is the key ingredient keeping a dream engine from being realized, rather than the asymmetry of the Onsager matrix. Thus, the BSC claim~\cite{Benenti} based on the Onsager asymmetry should be understood as a misleading result caused by overlooking the importance of the intrinsically imposed boundary condition.

\begin{acknowledgments}
 This research was supported by the NRF Grant No.~2017R1D1A1B06035497 (HP), No.~2017R1D1A1B03030872 (JU), and the KIAS individual Grants No.~PG013604 (HP), PG074001 (JMP), QP064902 (JSL) at Korea Institute for Advanced Study.

\end{acknowledgments}

\appendix

\section{Two-dimensional case}

Consider the equation of motion, Eq.~\eqref{eq:main_eq}, in the two-dimensional space with
\begin{align}
&\textsf{F}_\textrm{nc}= \left( \begin{array}{cc}
0 & \epsilon  \\
\delta & 0  \\
\end{array} \right),
\Gamma = \left( \begin{array}{cc}
\gamma & 0 \\
0 & \gamma \\
\end{array} \right), \textrm{ and }~
\xi =  \left( \begin{array}{c}
\xi_x  \\
\xi_y \\
\end{array} \right), \label{eq:matrix2}
\end{align}
where $\xi_i$ $(i=x,y)$ is a white Gaussian noise satisfying $\langle \xi_i (t) \xi_i  (t^\prime) \rangle =2\gamma T_i \delta(t-t^\prime)$ in the Boltzmann unit and $\textbf{B}=B \hat{z}$ in the ${z}$ direction. The thermodynamic forces are defined as
\begin{align}
X_1 \equiv\delta-\epsilon~~~\textrm{and}~~~X_2\equiv 1/T_x - 1/T_y~~~(T_x<T_y)~, \label{eq:force2}
\end{align}
and the currents are
\begin{align}
J_1 \equiv \frac{\langle x v_y \rangle_\textrm{s}}{T_x},~J_2  \equiv q_y=\frac{\gamma}{m}\left(T_y-m\langle v_y^2\rangle_\textrm{s}\right)~, \label{eq:J1J2a}
\end{align}
where $q_y$ is the heat current out of the high-temperature reservoir  and
the work current is given by $w=-J_1X_1T_x$.

In order to express the equation of motion in a multivariate Ornstein-Uhlenbeck form in Eq.\eqref{eqA:OU},
we introduce a state vector $\textbf{z} =(x,y,v_x,v_y)^{\textsf{T}}$ and a noise vector $d \Xi(t) = (d\Xi_1(t), d\Xi_2 (t), d\Xi_3 (t), d\Xi_4 (t))^{\textsf{T}}$ with $\langle d\Xi (t)\rangle =0$ and $\langle d\Xi(t) d\Xi^{\textsf{T}} (t)   \rangle = 2\textsf{D} dt$,
with
\begin{align}
\textsf{A}= \frac{1}{m} \left( \begin{array}{cccc}
0 & 0 & -m & 0  \\
0 & 0 & 0 & -m \\
k & -\epsilon & \gamma & -B  \\
-\delta & k & B & \gamma \\
\end{array} \right)~,~~
\textsf{D}= \frac{1}{m^2} \left( \begin{array}{cccc}
0 & 0 & 0 & 0  \\
0 & 0 & 0 & 0 \\
0 & 0 & \gamma T_x & 0  \\
0 & 0 & 0 & \gamma T_y \\
\end{array} \right)~. \label{eq:matrix22}
\end{align}

The covariant matrix $\Sigma$ satisfies Eq.~\eqref{eqA:OUsln} in the steady state and its expansion
near the equilibrium  $(\delta=\epsilon, T_y=T_x)$ can be obtained through Eqs.~\eqref{eq:OUper1} and \eqref{eq:OUper2}
with $\textsf{A}_0= \textsf{A}|_{\delta=\epsilon}$,  $\textsf{D}_0=\textsf{D}|_{T_y= T_x}$,
\begin{align}
\textsf{A}_1= \frac{1}{m} \left( \begin{array}{cccc}
0 & 0 & 0 & 0  \\
0 & 0 & 0 & 0 \\
0 & 0 & 0 & 0  \\
-1 & 0 & 0 & 0 \\
\end{array} \right),~\textrm{and}~~
\textsf{D}_2= \frac{1}{m^2} \left( \begin{array}{cccc}
0 & 0 & 0 & 0  \\
0 & 0 & 0 & 0 \\
0 & 0 & 0 & 0  \\
0 & 0 & 0 & \gamma T_x^2 \\
\end{array} \right)~. \label{eq:matrix22}
\end{align}

It is simple to find $\Sigma_0$ from Eq.~\eqref{eq:OUper2} as
\begin{align}
\Sigma_0=  T_x\left( \begin{array}{cccc}
{k}/K & \epsilon/K & 0 & 0  \\
\epsilon/K & k/K & 0 & 0 \\
0 & 0 & 1/m & 0  \\
0 & 0 & 0 & 1/m \\
\end{array} \right)~,~
\end{align}
with $K=k^2-\epsilon^2$. The stability condition is given by $K>0$.
We can also find $\Sigma_1$ and $\Sigma_2$ as well.

From Eqs.~\eqref{eq:Onsager} and \eqref{eq:J1J2}, the Onsager matrix $\textsf{L}$
is given as
\begin{align}
\textsf{L}=  \left( \begin{array}{cc}
[\Sigma_1]_{14}/T_x & [\Sigma_2]_{14}/T_x  \\
-\gamma [\Sigma_1]_{44} & \gamma \left(T_x^2/m-[\Sigma_2]_{44}\right) \\
\end{array} \right)~,\label{eq:OnsagerL2}
\end{align}
and finally we get
\begin{align}
&L_{11}=\frac{B^2+\gamma^2}{\gamma\mathcal{C}}~,~~ L_{22}=\frac{\gamma T_x^2(kB^2+m\epsilon^2)}{m\mathcal{C}}~, \nonumber\\
&L_{12}=\frac{-\epsilon\gamma T_x}{\mathcal{C}}=L_{21}~,~\textrm{with}~\mathcal{C}=2(kB^2+m\epsilon^2+k\gamma^2)~. \label{eq:L_2}
\end{align}
As seen in Eq.~\eqref{eq:L_2}, the Onsager matrix is an even function of the magnetic field $B$, thus
is symmetric $(\textsf{L}=\textsf{L}^\textsf{T}, s=1)$ like in other two-terminal particle transport systems~\cite{Buttiker, Brandner}. We also note that
$\mathcal{L}=4\textrm{det}(\textsf{L})=2B^2T_x^2/(m\mathcal{C})>0$ (no tight-binding), implying that the reversible process ($\langle \dot{S}_\textrm{tot}\rangle_\textrm{s}=0$) is impossible, thus the efficiency $\eta$ cannot reach the Carnot efficiency $\eta_\textrm{C}$
for non-zero $B$.

\section{Cyclic engine}

We consider a cyclic engine with time period $\tau$ as follows. An engine system is in contact with multiple heat reservoirs with temperature $T_i(t)$ varying periodically in time $t$ as $T_i(t+\tau) =T_i(t)$.
We assume that the system is described by a Langevin dynamics.
The average heat energy $\langle Q_i\rangle $ out of the $i$-th reservoir during one period is given by
\begin{align}
\langle {Q}_i \rangle &\equiv\int_0^\tau dt \langle \dot{Q}_i\rangle
\leq \int_0^\tau dt |\langle \dot{Q}_i\rangle | \leq \int_0^\tau dt
\sqrt{\zeta_i} \sqrt{\langle \dot{S}_\textrm{tot}\rangle} \nonumber\\
&\leq \sqrt{\int_0^\tau dt \sqrt{\zeta_i}} \sqrt{\int_0^\tau dt \langle \dot{S}_\textrm{tot}\rangle}
%\equiv \sqrt{\chi} \sqrt{}
~, \label{eq:AppB1}
\end{align}
where Eq.~\eqref{eq:dechant-sasa} and the Cauchy-Schwarz inequality are applied.
Then, we get the inequality similar to Eq.~\eqref{eq:dechant-sasa} as
\begin{align}
\langle {Q}_i \rangle^2\leq \chi_i \langle \Delta S_\textrm{tot}\rangle~,
\end{align}
with a positive constant $\chi_i=\int_0^\tau dt \sqrt{\zeta_i}$ and
the total EP during one period $\langle\Delta S_\textrm{tot}\rangle$.

With the two (hot and cold) reservoirs with temperatures $T_h$ and $T_c$ respectively, we can easily find
%Using $\Delta S_\textrm{tot} = \Delta S_\textrm{sys} - Q_\textrm{h}/T_\textrm{h} - Q_\textrm{c}/T_\textrm{c}$, where $\Delta S_\textrm{sys}$ is averaged Shannon entropy change of the system and $Q_\textrm{c} \equiv \int_{\tau_2}^{\tau_3}  dt \langle \dot{Q} \rangle_t $, we have
\begin{align}
\langle W\rangle  \leq \frac{\chi_h \eta}{T_c} \left[ \eta_\textrm{C} -\eta +  \frac{T_c \langle \Delta S_\textrm{sys}\rangle - \langle\Delta E\rangle}{\langle Q_h\rangle}  \right] ,
\label{eq:bound_periodic}
\end{align}
where $\langle W\rangle$,  $\langle \Delta S_\textrm{sys}\rangle $, and $\langle \Delta E\rangle $ are the work production,
the Shannon entropy change,  and  the system energy change during one period, respectively. In the cyclic steady state with $\langle\Delta S_\textrm{sys}\rangle = \langle\Delta E \rangle= 0$, the work extraction is impossible at the Carnot efficiency, even though
it is possible in a transient state.

%\vfil\eject


\begin{thebibliography}{99}
\bibitem{Kittel} C. Kittel and H. Kroemer, \emph{Thermal Physics} Ch. 8, 2nd Ed. (W. H. Freeman and Company, 1980).

\bibitem{Benenti} G. Benenti, K. Saito, and G. Casati,
Phys. Rev. Lett. \textbf{106}, 230602 (2011).

\bibitem{Onsager} L. Onsager, Phys. Rev. \textbf{38}, 2265 (1931).
\bibitem{Casimir} H. B. G. Casimir, Rev. Mod. Phys. \textbf{17}, 343 (1945).

\bibitem{HKLee} H. K. Lee, S. Lahiri, and H. Park, Phys. Rev. E \textbf{96}, 022134 (2017).
\bibitem{Brandner} K. Brandner, K. Saito, and U. Seifert, Phys. Rev. Lett. \textbf{110}, 070603 (2013).

\bibitem{Allahverdyan1} A. E. Allahverdyan, K. V. Hovhannisyan, A. V. Melkikh, and S. G. Gevorkian,
%Carnot Cycle at Finite Power: Attainability of Maximal Efficiency,
Phys. Rev. Lett. \textbf{111}, 050601 (2013).

\bibitem{Karel} K. Proesmans and C. Van den Broeck,
%Onsager Coefficients in Periodically Driven Systems,
Phys. Rev. Lett. \textbf{115}, 090601 (2015).

\bibitem{Campisi} M. Campisi and R. Fazio,
%The power of a critical heat engine,
Nat. Commun. \textbf{7}, 11895 (2016).

\bibitem{Shiraish} N. Shiraishi, K. Saito, and H. Tasaki,
%Universal Trade-Off Relation between Power and Efficiency for Heat Engines,
Phys. Rev. Lett. \textbf{117}, 190601 (2016).

\bibitem{Polettini} M. Polettini and M. Esposito,
%Carnot efficiency at divergent power output,
EPL, \textbf{118}, 40003 (2017).

\bibitem{Holubec1} V. Holubec and A. Ryabov,
%Cycling Tames Power Fluctuations near Optimum Efficiency,
Phys. Rev. Lett. \textbf{121}, 120601 (2018).

\bibitem{Andreas} A. Dechant and S.-I. Sasa,
%Entropic bounds on currents in Langevin systems,
Phys. Rev. E \textbf{97}, 062101 (2018).

\bibitem{Pietzonka} P. Pietzonka and U. Seifert,
%Universal trade-off between power, efficiency and constancy in steady-state heat engines,
Phys. Rev. Lett. \textbf{120}, 190602  (2018).

\bibitem{JSLee1} J. S. Lee and H. Park,
%Carnot efficiency is reachable in an irreversible process.
Sci. Rep. \textbf{7}, 10725 (2017).

\bibitem{JSLee2} J. S. Lee, S. H. Lee, J. Um, and H. Park,
%Carnot efficiency and zero-entropy-production rate do not guarantee reversibility of a process,
J. Korean Phys. Soc. \textbf{75}, 948 (2019).


\bibitem{Buttiker} M. B\"{u}ttiker,
%Symmetry of electrical conduction,
IBM J. Res. Dev. \textbf{32}, 317 (1988).

\bibitem{Stark} J. Stark, K. Brandner, K. Saito, and U. Seifert,
%Classical Nernst engine,
Phys. Rev. Lett. \textbf{112}, 140601 (2014).

\bibitem{Brandner2} K. Brandner and U. Seifert,
%Multi-terminal thermoelectric transport in a magnetic field: bounds on Onsazger coefficients and efficiency,
New J. Phys. \textbf{15}, 105003 (2013).

\bibitem{exp1} One may take the x-axis dynamics also affected by either reservoir, still making a two-terminal
engine. The calculation is a bit more complicated, but the main conclusion does not change.

\bibitem{Crisanti} A. Crisanti, A. Puglisi, and D. Villamaina,
%Nonequilibrium and information: The role of cross correlations,
Phys. Rev. E \textbf{85}, 061127 (2012).

\bibitem{ParkJM} J.-M. Park, H.-M. Chun, and J. D. Noh,
%Efficiency at maximum power and efficiency fluctuations in a linear Brownian heat-engine model,
Phys. Rev. E \textbf{94}, 012127 (2016).
\bibitem{Filliger} R. Filliger and P. Reimann,
%Brownian Gyrator: A Minimal Heat Engine on the Nanoscale,
Phys. Rev. Lett. \textbf{99}, 230602 (2007).

\bibitem{Chiang} K.-H. Chiang, C.-L. Lee, P.-Y. Lai, and Y.-F. Chen,
%Electrical autonomous Brownian gyrator,
Phys. Rev. E \textbf{96}, 032123 (2017).
\bibitem{LeeKwon} S. Lee and C. Kwon, Phys. Rev. E \textbf{99}, 052142 (2019).
\bibitem{Chun1} H.-M. Chun, L. P. Fischer, and U. Seifert, Phys. Rev. E \textbf{99}, 042128 (2019).
%\bibitem{Chun2} H.-M. Chun, J. Um, and H. Park (unpublished).

\bibitem{Sekimoto} K. Sekimoto, Prog. Theor. Phys. \textbf{130}, 17 (1998).
%\bibitem{JMPark}  J.-M. Park and J. D. Noh, Phys. Rev. E \textbf{94}, 012127 (2016).

\bibitem{vdBroeck} C. Van den Broeck, Phys.Rev. Lett. \textbf{95}, 190602 (2005).
\bibitem{Gardiner} C. Gardiner, \emph{Stochastic Methods} Ch. 4, 4th Ed. (Springer-Verlag, Berlin, 2009).

\bibitem{LeePark2} J. S. Lee, J.-M. Park, and H. Park (unpublished).


\end{thebibliography}
\end{document}